\documentclass[conference]{IEEEtran}
\IEEEoverridecommandlockouts

\usepackage{cite}
\usepackage{amsmath,amssymb,amsfonts}
\usepackage{algorithmic}
\usepackage{graphicx}
\usepackage{textcomp}
\usepackage{xcolor}
\usepackage{subcaption}
\def\BibTeX{{\rm B\kern-.05em{\sc i\kern-.025em b}\kern-.08em
    T\kern-.1667em\lower.7ex\hbox{E}\kern-.125emX}}
\begin{document}

\title{AirCompSim: A Discrete Event Simulator for \\Air Computing}

\author{\IEEEauthorblockN{Baris Yamansavascilar}
\IEEEauthorblockA{\textit{Department of Computer Engineering} \\
\textit{Bogazici University}\\
Istanbul, Turkiye \\
baris.yamanasavascilar@std.bogazici.edu.tr}
\and
\IEEEauthorblockN{Atay Ozgovde}
\IEEEauthorblockA{\textit{Department of Computer Engineering} \\
\textit{Bogazici University}\\
Istanbul, Turkiye \\
ozgovde@bogazici.edu.tr}
\and
\IEEEauthorblockN{Cem Ersoy}
\IEEEauthorblockA{\textit{Department of Computer Engineering} \\
\textit{Bogazici University}\\
Istanbul, Turkiye \\
ersoy@bogazici.edu.tr}
}

\maketitle

\begin{abstract}

Air components, including UAVs, planes, balloons, and satellites have been widely utilized since the fixed capacity of ground infrastructure cannot meet the dynamic load of the users. However, since those air components should be coordinated in order to achieve the desired quality of service, several next-generation paradigms have been defined including air computing. Nevertheless, even though many studies and open research issues exist for air computing, there are limited test environments that cannot satisfy the performance evaluation requirements of the dynamic environment. Therefore, in this study, we introduce our discrete event simulator, AirCompSim, which fulfills an air computing environment considering dynamically changing requirements, loads, and capacities through its modular structure. To show its capabilities, a dynamic
capacity enhancement scenario is used for investigating the effect of the number of users, UAVs, and requirements of different application types on the average task success rate, service time, and server utilization. The results demonstrate that AirCompSim can be used for experiments in air computing.

\end{abstract}

\begin{IEEEkeywords}
Air Computing, UAVs, non-terrestrial, simulator
\end{IEEEkeywords}

\section{Introduction}

The extensive utilization of smart devices by end-users results in various application types that have different goals. Therefore, providing a necessary quality of service (QoS) based on the corresponding service level agreement (SLA) requirements is currently challenging for the existing network paradigms. Even though multi-access edge computing (MEC) \cite{wang2018power} has been widely used for this purpose, its static deployment in which the capacity of the edge server(s) cannot be changed based on the dynamic needs would cause an issue to meet the requests of the user tasks. Thus, unmanned aerial vehicles (UAVs) and similar aerial units have recently been used by many studies in order to enhance the performance of the network environment in terms of QoS \cite{zhang2024enhancing, hao2024joint, shi2024task}.

\begin{figure}[t]
\centering
\frame{\includegraphics[scale=0.055]{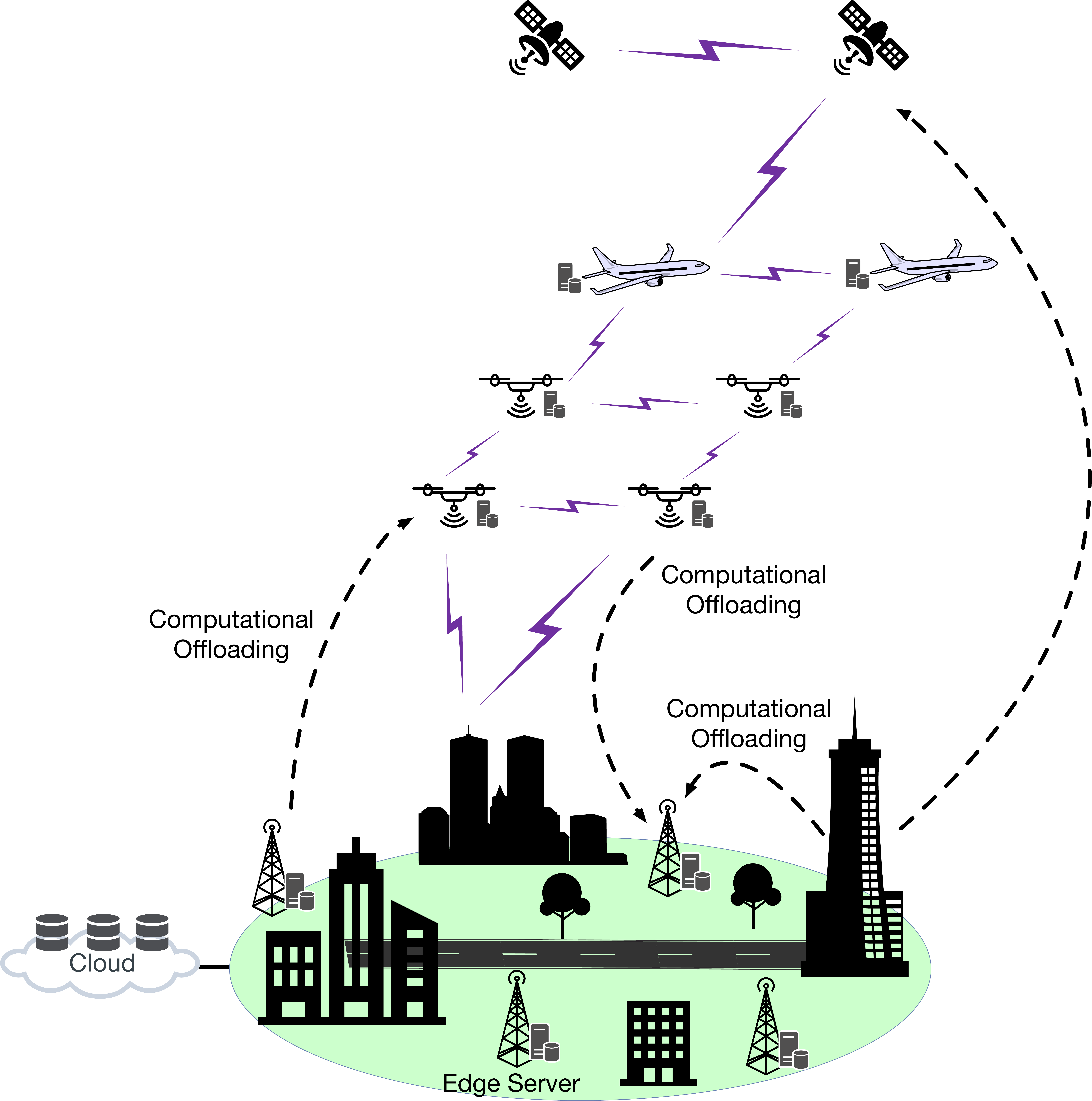}}
\caption{An air computing environment.}
\label{AirComputing}
\end{figure}

Considering varied deployment cases of UAVs and related air vehicles to alleviate the restricted 2D networking environment, we recently proposed air computing which is a next-generation computational paradigm as a result of the evolution of MEC \cite{yamansavascilar2024air}. In this regard, air computing represents a dynamic and responsive computation environment for all spectrum of applications through different air platforms. Hence, in addition to the ground layer, there are three air platforms in air computing as shown in Figure \ref{AirComputing}. These are low altitude platform (LAP) in which UAVs are deployed, high altitude platform (HAP) where airplanes and balloons operate, and low earth orbit (LEO) in which satellites are run. Each of these air platforms can provide different opportunities in terms of latency, data rate, computational capability, coverage, and mobility to the corresponding applications using the benefits of 3D networking. The joint operation of those air platforms in a dynamic environment to meet different requirements are still investigated.

Even though air platforms, especially UAVs, have recently been used by many studies for varied goals such as task offloading, content caching, and trajectory planning, there are few simulators that consider these scenarios in the literature. Moreover, they do not allow mobile users, applicable UAV policies, and the definition of different application types, which may have different SLA requirements, such as maximum tolerable delay. In most of the cases, existing simulators focus on the flight controls, visualization and physics \cite{ebeid2018survey}. In \cite{shah2018airsim}, Shah et al. proposed AirSim which is built on Unreal Engine and focused on realistic flying simulation of aerial vehicles. Their simulator includes a physics engine for real-time hardware-in-the-loop simulations. D'Urso et al. represented an integrated simulator for UAVs for the realistic simulations in \cite{d2019integrated}. Therefore, they developed a software middleware that coordinates the tools including Gazebo \cite{koenig2004design}, ArduCopter \cite{ArduPilot}, and ns-3 \cite{riley2010ns} to work together. Similarly, Baidya et al. proposed a middleware for the realistic UAV simulation in \cite{baidya2018flynetsim}. They focused on the network dynamics and modeling UAV operations so that they interfaced ArduPilot \cite{ArduPilot} and ns-3 tools through their simulator. To the best of our knowledge, none of the existing simulators consider application performance in terms of task completion, user mobility, and UAV flying policy based on the system load and computational resources.

In this study, we introduce a new simulator, namely AirCompSim, which ensures a discrete event simulation environment to conduct complex air computing experiments based on varied scenarios \cite{AirCompSim}. AirCompSim achieves that by providing a modular structure. Therefore, it is straightforward to alter the corresponding parameters, add new scenarios including dynamic events such as server failure, and create new flying policies for UAVs. Moreover, it also provides an event mechanism for deep reinforcement learning (DRL) so that smart and sophisticated UAV policies such as where to fly and when to fly can be investigated. Furthermore, since the essential results are logged and stored, it provides an easy mechanism to plot the related results after the experiments. To demonstrate the capabilities of AirCompSim, we present a dynamic capacity enhancement scenario and evaluate the performance of different settings in an air computing environment. The main contributions of our paper are as follows.

\begin{itemize}

\item We present AirCompSim, which provides air computing simulations considering different user/application profiles, edge servers, and air vehicles, especially UAVs. Therefore, to the best of our knowledge, AirCompSim has unique features regarding the existing simulators.

\item AirCompSim supports dynamic scenarios through its scenario module in which deployed entities would be failed, or existing policies can be updated. Moreover, it provides a DRL-based event mechanism so that a researcher can implement a novel flying or offloading policy in this manner. Furthermore, based on the event mechanism, new modules based on different methods can easily be incorporated in AirCompSim.

\end{itemize}

The rest of this paper is organized as follows. In Section II, we elaborate on the background of air computing. Next, we represent the architecture of AirCompSim in Section III. In Section IV, we explain the dynamic capacity enhancement use-case to implement in our simulator and afterwards we show the experimental results. Finally, we conclude our study in Section V.




\section{Air Computing Background}

The essential idea of edge computing is that offloading the computation-intensive tasks from end devices to the corresponding edge servers since battery and CPU limitations cannot allow the local execution. Therefore, users decide where to offload and when to offload the corresponding tasks if there are multiple edge servers nearby. However, since the delay requirements change for the mission-critical applications, and mobile devices including tablets and smartwatches proliferate, traditional edge computing based on terrestrial resources would be insufficient to meet the suitable computing capacity. As a remedy, air-based computational resources have recently been proposed to enhance computational capacity by augmenting 3D networking opportunities.

The most popular implementation of this paradigm is UAV-assisted edge computing  since UAVs provide flexibility in terms of flying, and lower latency since they are close to the ground. However, other air vehicles including airplanes, balloons, and LEOs are also used for this purpose. Thus, air computing includes all air vehicles in order to enhance the edge computing paradigm. As a result, we believe that air computing is the evolution of edge computing through air vehicles. To this end, air computing consists of three air platforms including LAP, HAP, and LEO as shown in Figure \ref{AirComputing}. Each platform provides different features regarding the requirements of the underlying environment. 
 
\subsection{LAP}

The main deployment of LAP is on urban areas in which the existing infrastructure is built well and therefore meets the QoS of user applications. Since the operational altitude of the corresponding air vehicles in LAP, which are UAVs, is below 10 km, the propagation delay would change between 10 - 30 $\mu$s. Moreover, since they can provide Line of Sight (LoS), connectivity, service provision, and low latency can be ensured seamlessly.



\subsection{HAP}

Air vehicles in HAP can be used on urban and suburban areas since they can fly at high altitudes between 10 - 30 km. Therefore, their propagation delay changes between 50 - 85 $\mu$s. Moreover, channel and weather conditions also affect  communication quality. Thus, they cannot be used for mission-critical applications whose delay tolerance is low.

The essential use case for HAP is for regional coverage in which airplanes and balloons can be deployed as management nodes for UAVs and terrestrial servers. Moreover, they can be used as computational resources if the SLA requirements of the corresponding tasks would not be violated in terms of  delay. 



\subsection{LEO}

LEO platform consists of satellites whose altitude changes between 160 - 2000 km. Because of this altitude range, its propagation delay would be between 1.5-3 ms which is not suitable for very low latency applications. However, they can carry out edge computing solutions through task offloading as either used as a relay node or using their limited onboard capacity. Moreover, they are also used to access cloud computing solutions.


\section{AirCompSim Architecture}

\begin{figure}[t]
\centering
\includegraphics[scale=0.35]{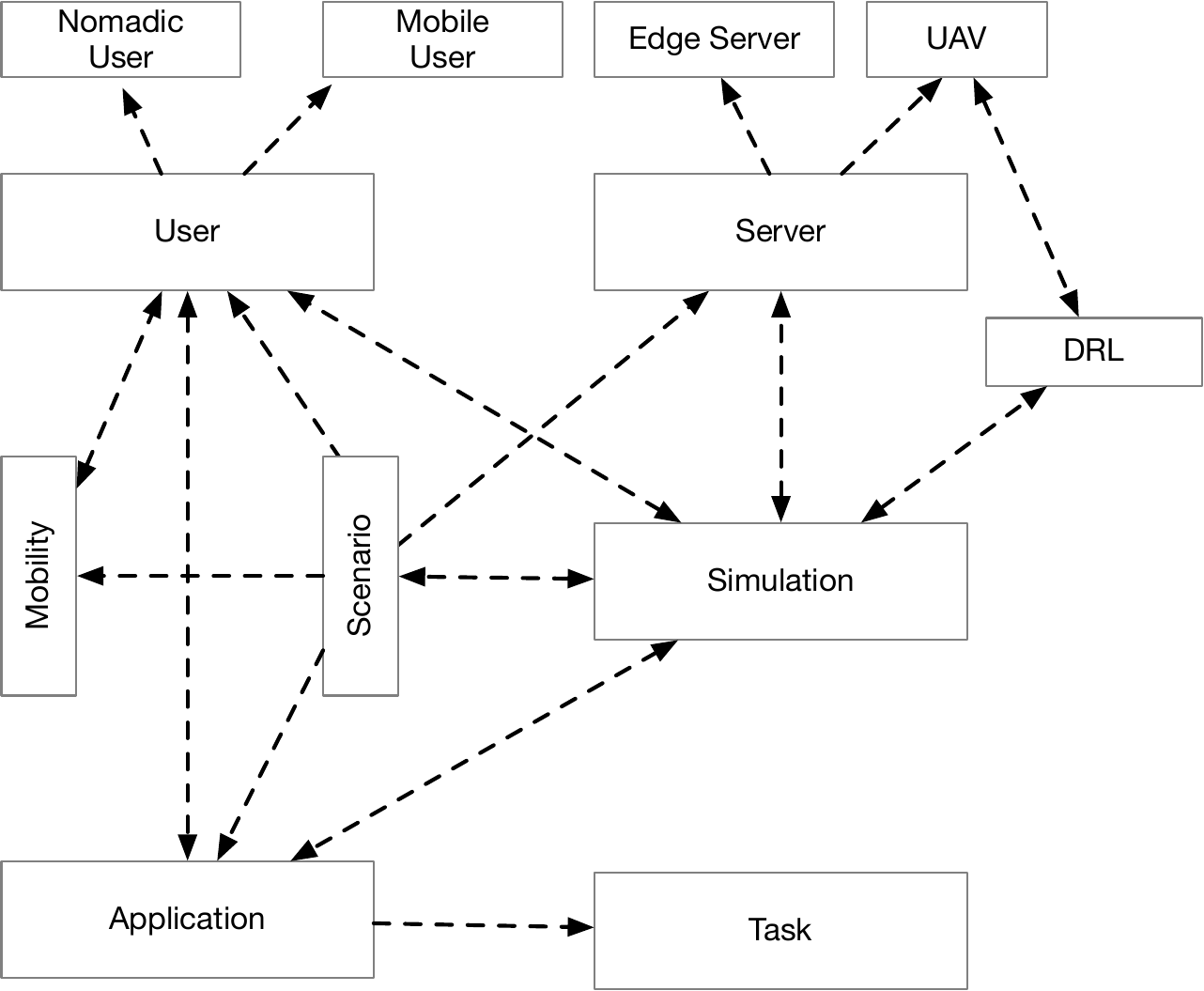}
\caption{Relationship of the AirCompSim modules.}
\label{AirCompSim}
\end{figure}

AirCompSim provides a modular architecture in which each module can interact with each other through the corresponding instances and interfaces. Moreover, the modular architecture based on the discrete event system ensures that a new module can be easily incorporated into AirCompSim. To this end, AirCompSim has five core modules namely Simulator, Server, User, Application, and Task. The relationship of these five modules is shown in Figure \ref{AirCompSim}. 

\subsection{Simulation Module}

The simulation module is the main module of AirCompSim. It is responsible for handling the events in a simulation via the event queue and creating the corresponding user types, user mobility, and UAV policies. Moreover it provides a detailed logging mechanism including the events, task offloading, UAV flying, and task processing so that it allows a detailed debugging option for the researchers. Furthermore, it saves the results of the repeated simulations as a \textit{csv} file therefore after a detailed experiment it is straightforward to process the results using various libraries such as \textit{pandas} \cite{mckinney2011pandas}.

\subsection{Server Module}

The server module defines a generic class for the entities having server features. To this end, this module is used as a parent class for edge servers, UAVs, and the cloud server. Note that in the current version of AirCompSim \cite{AirCompSim}, UAVs are the only flying vehicles in the simulator. However, it is easy to add additional flying servers with different altitudes based on the server module. This module also provides the utilization, processing delay for the given tasks, and earliest idle time based on the queueing. A server module-based entity has a computational capacity to serve the offloaded tasks based on the $M/M/1$ queueing model.

\subsubsection{Edge Server}

In our simulator, edge servers are located in LANs on the ground. They have a coverage radius in which users can connect to them to offload their tasks. Their capacity and location are fixed throughout an experiment. Even though the corresponding capacity can be set by researchers in the simulation module, in the default settings of AirCompSim, the capacity of an edge server is less than that of a cloud server and higher than that of a UAV.

\subsubsection{UAV}

We consider UAVs as flying edge servers with less capacity in our simulator. Therefore, various flying policies can be applied in AirCompSim in order to observe the effects of different proposals on task success rate, utilization, and service delay. 
In the default settings of AirCompSim, UAVs fly to the areas that have already been covered by an edge server in order to enhance the corresponding capacity.

\subsubsection{Cloud Server}

A cloud server is the most powerful entity in terms of computational capacity in an air computing environment. However, since it is accessed through the WAN, it is prone to higher network delay than LAN and MAN. Hence, it is critical to select a cloud server for task offloading when the corresponding maximum tolerable delay is high. In the default settings of AirCompSim, a cloud server is selected by a user if there is neither an edge server nor a UAV to connect.

\subsection{User Module}

The user module supports different user types, including mobile users, nomadic users, and users in the air. In the current implementation of AirCompSim there are mobile users and nomadic users. The main difference between them is their mobility pattern. A user can connect to multiple servers simultaneously if it is in the coverage of them. In the default mode of AirCompSim, if a user is connected to multiple servers, such as a UAV and an edge server, the task is offloaded to one of them, which has a lower queueing delay. Here, we assume that users are informed by their connected servers in a separate channel. Note that this default behavior can be changed by a researcher to seek efficient offloading decisions.


\subsection{Application Module}

Each user in the simulator runs at least one application, which randomly produces a computation-intensive task based on the application type. Therefore, each task based on an application type consists of a size, required computational units for processing, arrival rate, and maximum tolerable delay tuple. Each task is successfully completed if the total delay, which includes network, processing, and queueing delays, is lower than or equal to the maximum tolerable delay. The performance of the applied policies is evaluated based on the total task success rate, which also indicates the success rate of the corresponding application. For this reason, the location of the users affects the performance since some areas in the environment may not have an infrastructure, such as edge servers, and other places may face congestion. 

\subsection{Mobility Module}

The mobility module provides user mobility and also heuristic UAV flying policies. Based on the research goals, a user mobility module or a flying policy for UAVs can be added to the simulator. By default, a random waypoint model runs for the mobility of each user in the environment. On the other hand, UAVs fly between areas where edge servers are deployed in the center based on capacity calculations. A capacity calculation for areas is carried out by including the number of user and their application profiles, and also the total capacity of edge servers. Afterwards, the required number of additional capacity and therefore the number of UAVs are computed. Finally, UAVs are deployed in the corresponding areas. Note that if an area has a higher need for UAVs, that area has a priority for UAV deployment considering a constraint in the number of UAVs.

\begin{figure}[t]
\centering
\includegraphics[scale=0.45]{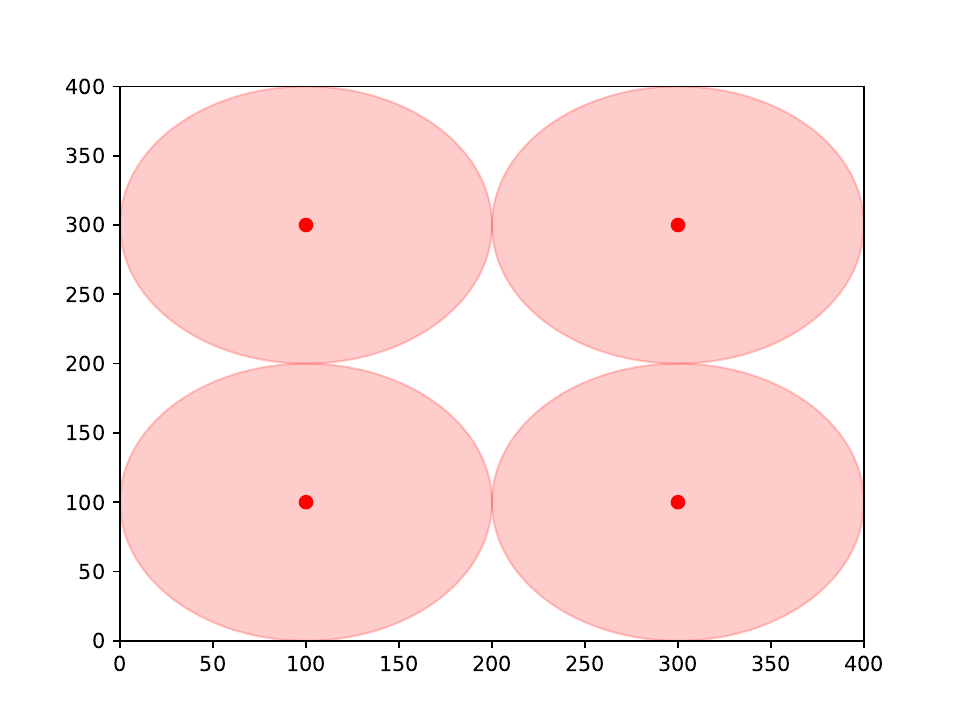}
\caption{Distribution and coverage of edge servers in the example scenario.}
\label{Edge-Radius}
\end{figure} 

\subsection{Scenario Module}

The scenario module includes the experimental setup which is essential for the performance evaluation. To this end, the number of users, edge servers, UAVs, the location of all kinds of servers, the user mobility model, and also the corresponding capacities of servers are first defined in this module. Moreover, if the experiment includes dynamic updates such as a server failure or sudden increase in capacity demand, those can also be defined in this module.

\subsection{DRL Module}

The DRL module consists of the implementation of value-based and policy-based DRL algorithms in addition to the event mechanism which provides the corresponding state information along with the rewards. The implementation of DRL algorithms is supported by PyTorch-based neural networks. Note that since DRL-based studies require more sophisticated settings, AirCompSim only provides the skeleton that can be adapted to the related scenarios by researchers.



\section{Use-Case: Dynamic Capacity Enhancement}

In order to demonstrate the capabilities of AirCompSim, we developed a dynamic capacity enhancement scenario in which we investigated the effect of the number of UAVs and users on the task success rate, edge utilization, UAV utilization, service time, and the ratio of the offloaded tasks between edge, cloud, and UAV. Moreover, we also evaluated their effect on different application types in terms of task success rate, which is, to the best of our knowledge, provided by none of the existing simulators. 

\subsection{Scenario}

In our scenario, we have a $400 \times 400 m^2$ environment. We divided the environment into four equal areas and deployed four edge servers in the center of them, as shown in Figure \ref{Edge-Radius}. On the other hand, each user in the environment is mobile and runs four different application types. Since we run the default mode of AirCompSim, deployed UAVs can only move between four edge server areas. Therefore, as depicted in Figure \ref{Edge-Radius}, there are areas in the environment that cannot be covered by any server. Users in those areas offload their tasks to the cloud server. To control the variance of results, we repeated our experiments 50 times. Throughout the experiments, we used Python 3.10. The corresponding simulation and application parameters are given in Tables \ref{SimParameters} and \ref{AppParameters}, respectively.

\begin{table}[!t]
\caption{Simulation Parameters}
\label{SimParameters}
\centering
\begin{tabular}{|l|l|}
\hline

\textbf{Parameter} & \textbf{Value}\\
\hline
\hline
Size of a task & 500 Kb\\
\hline
Capacity of a serving UAV & 1000 cmp. units/sec\\
\hline
Edge Server Radius & 100 m\\
\hline
UAV Radius & 100 m\\
\hline
Data Rate & 100 Mbps\\
\hline
$X_{max}$ & 400 m \\
\hline
$Y_{max}$ & 400 m \\
\hline
Altitude of UAVs & 200 m \\
\hline
Simulation Time & 1000 sec \\
\hline
User Mobility Model & Random Waypoint \\
\hline
\end{tabular}
\end{table}

\begin{table}[!t]
\caption{Application Parameters}
\label{AppParameters}
\centering
\begin{tabular}{|l|l|l|l|}
\hline

\textbf{Application} & \textbf{\shortstack{Arrival \\ Rate} } & \textbf{\shortstack{Comp. \\ Load}} & \textbf{\shortstack{Max. \\ Tolerable\\ Delay}}\\
\hline
\hline
Entertainment & 10 sec/task & 100 units & 0.3 sec\\
\hline
Multimedia & 10 sec/task & 100 units & 3 sec\\
\hline
Rendering & 20 sec/task & 200 units & 1 sec\\
\hline
Image Classification & 20 sec/task & 600 units& 1 sec\\
\hline
\end{tabular}
\end{table}

\subsection{Performance Evaluation}

\begin{figure}[t]
\centering
\includegraphics[scale=0.45]{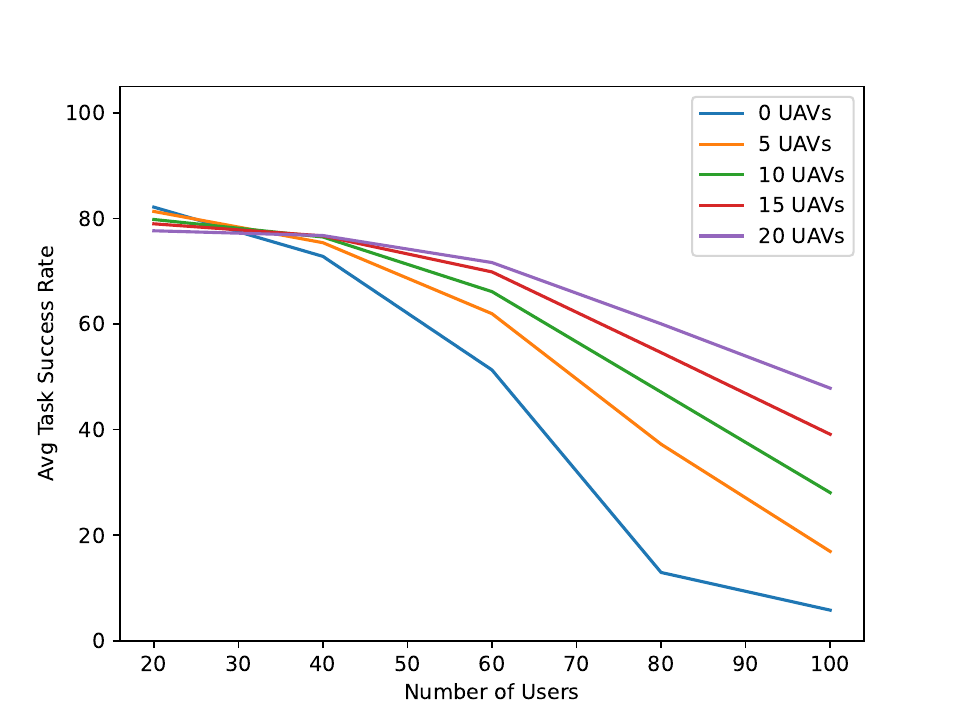}
\caption{Average task success rate.}
\label{Overall}
\end{figure} 

\begin{figure}[t]
\centering
\includegraphics[scale=0.45]{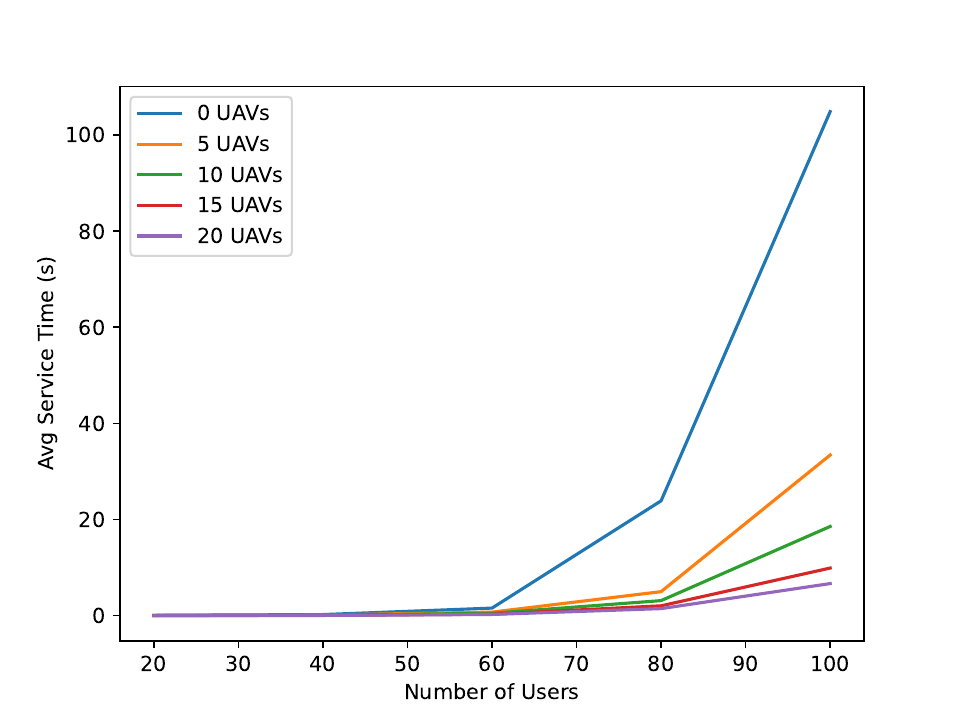}
\caption{Average service time.}
\label{ServiceTime}
\end{figure}

We first evaluated the effect of the number of users and UAVs on the average task success rate in the environment. As shown in Figure \ref{Overall}, the task success rate decreases when the number of users increases since the load in the system affects the utilization of deployed servers. However, this effect would be mitigated by deploying UAVs since they can fly dynamically between edge server areas to enhance the existing capacity. As a result of this, the task success rate is improved when more UAVs are deployed in the system. On the other hand, regarding the simulation and application parameters, the task success rate would be at most 80\% in this setting. The first reason for this result is the infrastructure-less areas as shown in Figure \ref{Edge-Radius}. Since UAVs move to the edge server areas in the default policy of the simulator, users in infrastructure-less areas can offload their tasks only to the cloud which causes high WAN delay. Therefore, application tasks, except the multimedia whose delay tolerance is 3 seconds, would not be completed successfully when they are offloaded to the cloud. The second reason for the task success rate results is related to queueing delay at servers based on their capacity and arrival rate of the offloaded tasks. Hence, the average service time in the system increases exponentially as shown in Figure \ref{ServiceTime}. In these results, when no UAVs are deployed, offloaded tasks undergo longer service time which affects the task success rate heavily. However, when we deploy UAVs, the service time decreases since users can offload the corresponding tasks to UAVs based on the existing queueing delays of all types of servers.

\begin{figure}[!t]
\centering
\begin{subfigure}{0.35\textwidth}
\includegraphics[width=\linewidth]{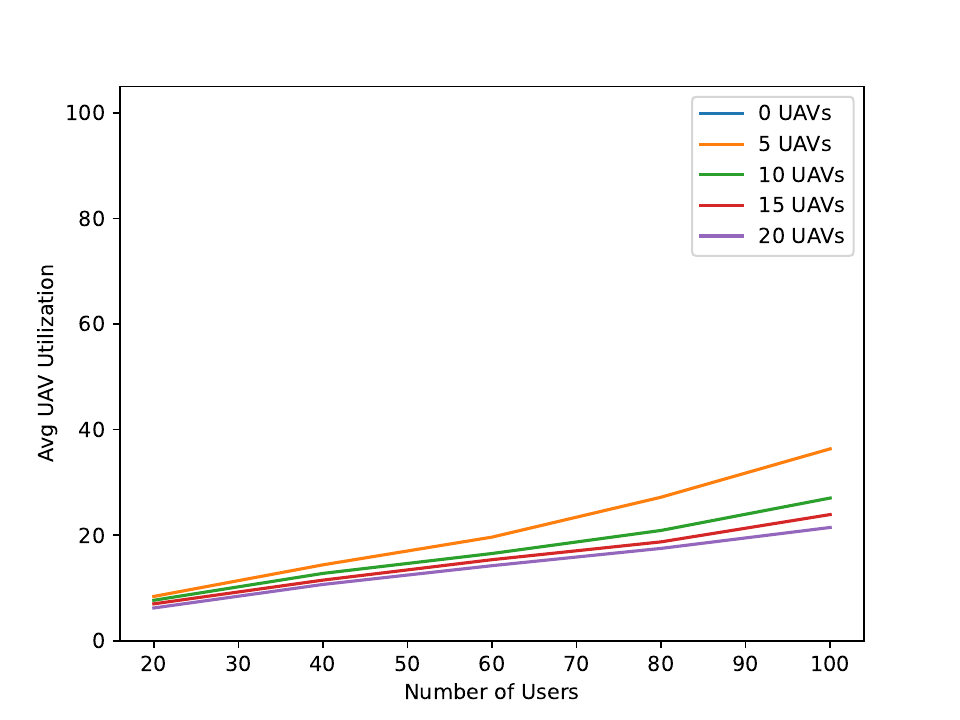}
\caption{UAV Utilization} \label{uti-1}
\end{subfigure}
\begin{subfigure}{0.35\textwidth}
\includegraphics[width=\linewidth]{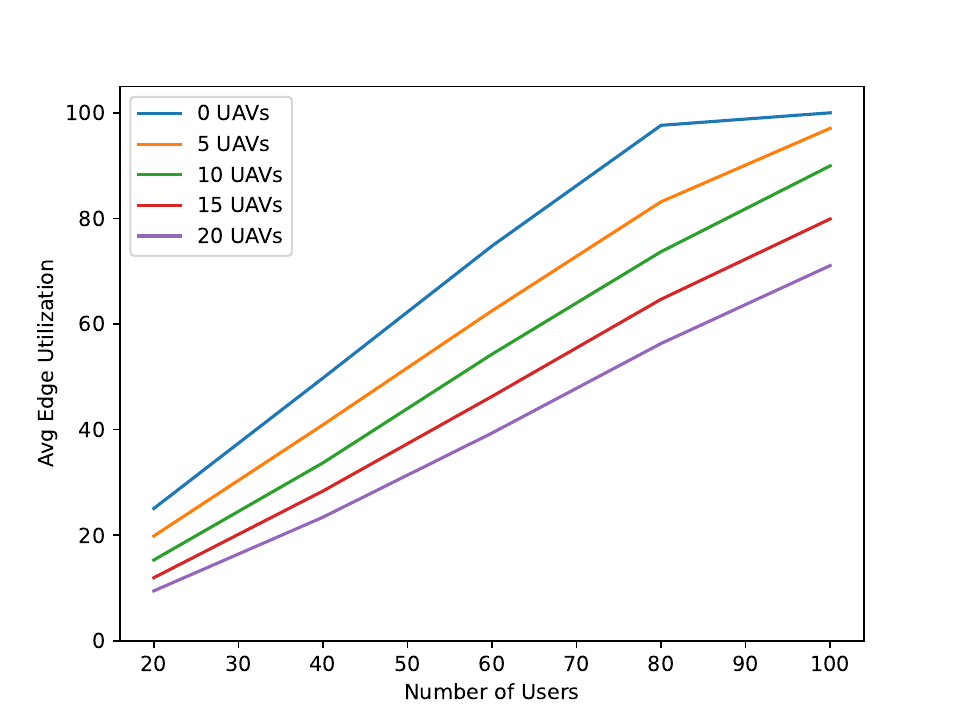}
\caption{Edge Utilization} \label{uti-2}
\end{subfigure}
\caption{Average UAV and Edge Utilization}
\label{utilization}
\end{figure}

Observing the utilization of the servers in the environment is also essential to analyze system resources. To this end, UAV and edge utilizations based on the number of users and UAVs are shown in Figure \ref{utilization}. When there is no UAV in the environment, edge servers are fully utilized while there are 80 and 100 users. On the other hand, for more UAV deployments, the edge utilization decreases since users offload their tasks to the available UAVs. Considering the UAV utilization as shown in \ref{uti-1}, when there are fewer UAVs in the environment, their average utilization is higher as in this case the total capacity is more limited.  


\begin{figure*}[!t]
\begin{subfigure}{0.24\textwidth}
\includegraphics[width=\linewidth]{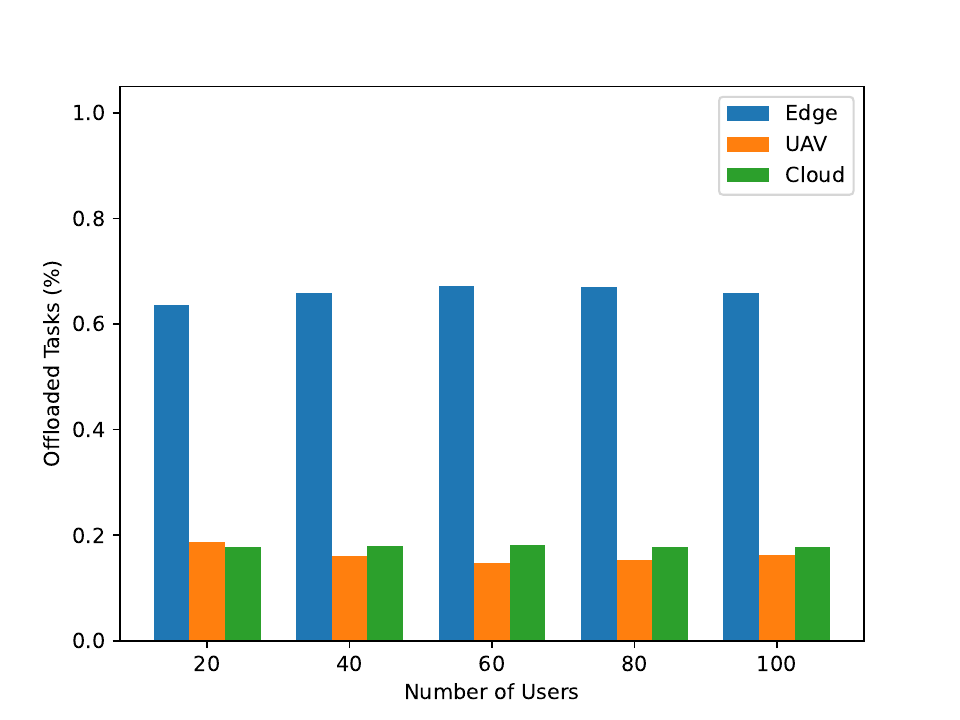}
\caption{5 UAVs} \label{5-uavs}
\end{subfigure}
\begin{subfigure}{0.24\textwidth}
\includegraphics[width=\linewidth]{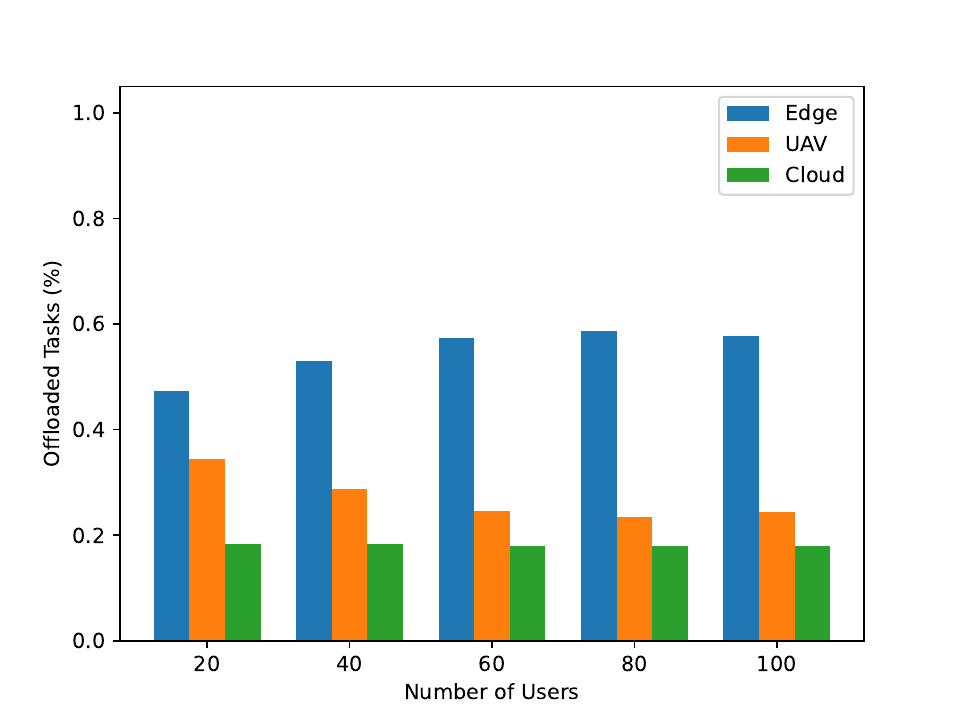}
\caption{10 UAVs} \label{10-uavs}
\end{subfigure}
\begin{subfigure}{0.24\textwidth}
\includegraphics[width=\linewidth]{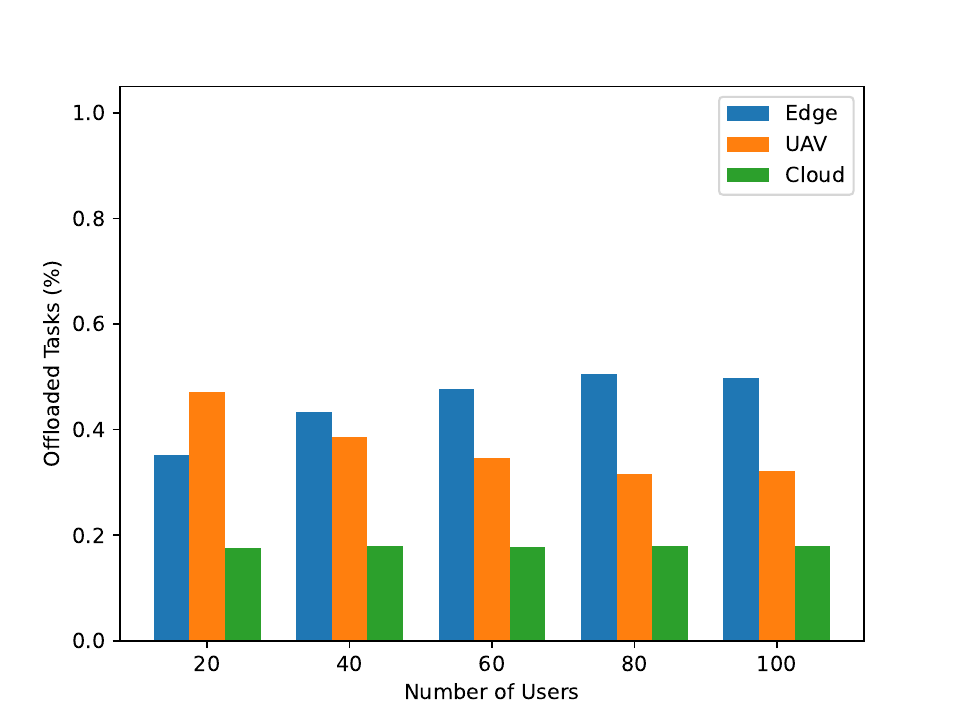}
\caption{15 UAVs} \label{15-uavs}
\end{subfigure}
\begin{subfigure}{0.24\textwidth}
\includegraphics[width=\linewidth]{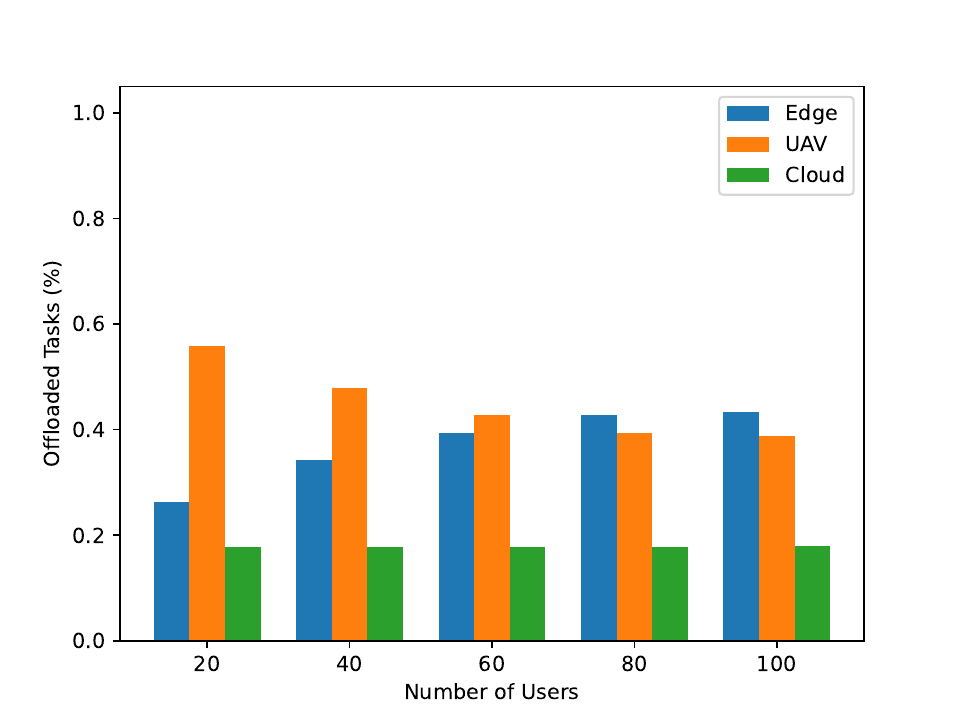}
\caption{20 UAVs} \label{20-uavs}
\end{subfigure}
\caption{Effect of deployed UAVs on offloading}
\label{offloaded}
\end{figure*}

Another effect of the deployment of UAVs is shown in Figure \ref{offloaded} which manifests the distribution of offloaded tasks between edge, cloud, and UAVs. Initially, the largest portion of tasks are offloaded to edge servers since most of the environment is covered by them and there are a small number of UAVs. However, for each increment of the number of UAVs, the corresponding offloading distribution shifts to UAVs. The main reason for this case is that users select the less congested server for their tasks which is the default task offloading policy in AirCompSim. On the other hand, the offloading percentage for the cloud server would not change since the infrastructure-less area is fixed in the environment.

\begin{figure}[!t]
\centering
\begin{subfigure}{0.23\textwidth}
\includegraphics[width=\linewidth]{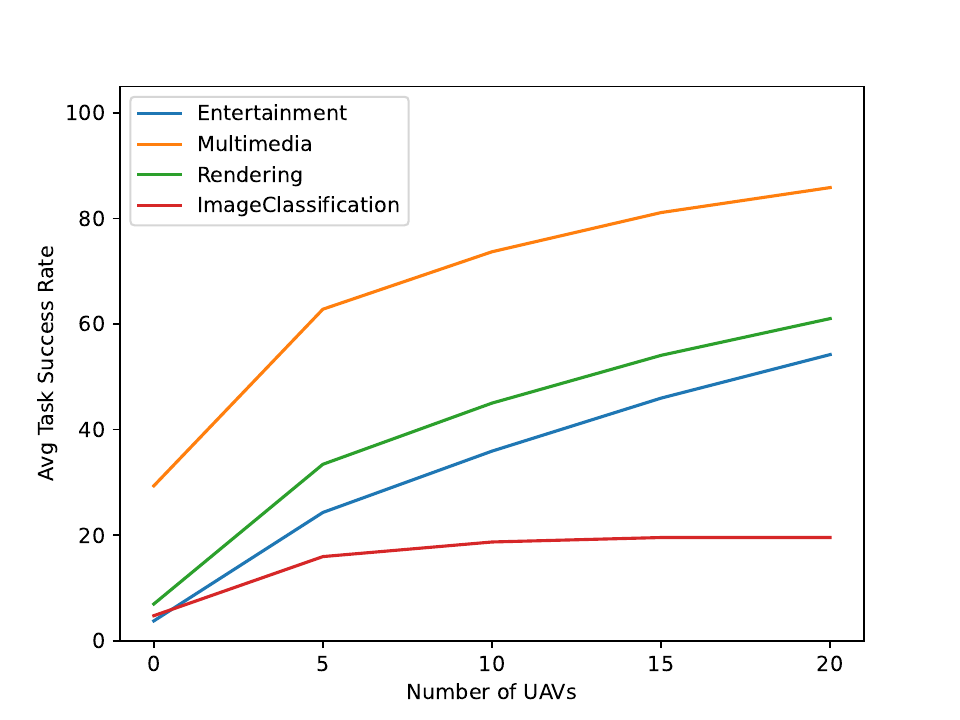}
\caption{80 users} \label{app-1}
\end{subfigure}
\begin{subfigure}{0.23\textwidth}
\includegraphics[width=\linewidth]{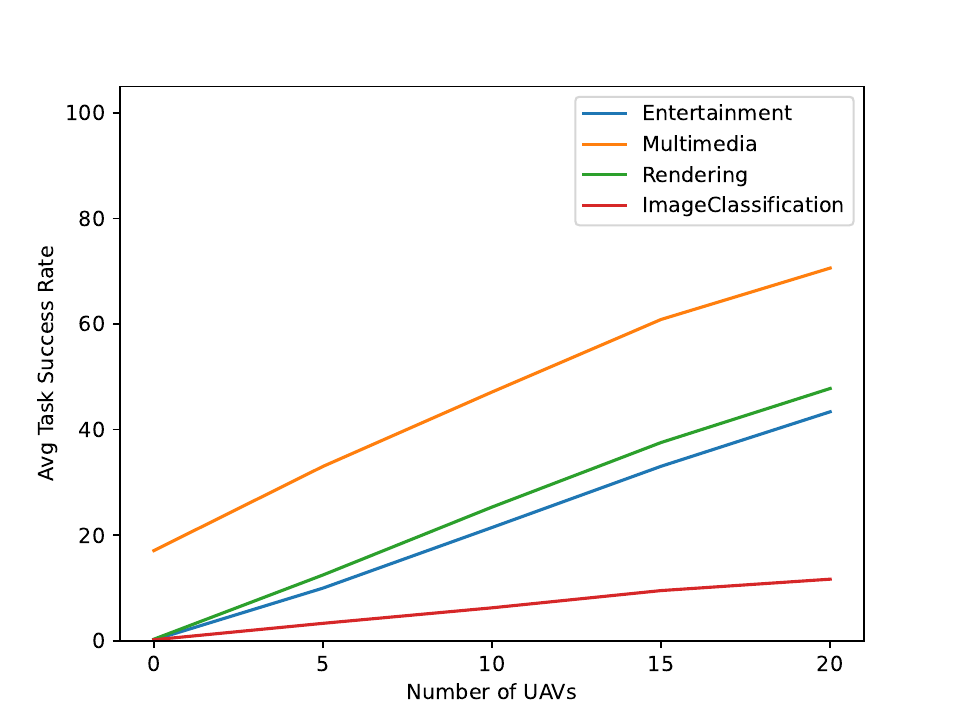}
\caption{100 users} \label{app-2}
\end{subfigure}
\caption{Task success rate of each application type}
\label{appSuccess}
\end{figure}

Finally, we investigated the average task success rate for each application, which is shown in Figure \ref{appSuccess}. Considering the application parameters given in Table \ref{AppParameters}, these are the expected results. The Entertainment application requires 100 computational units, which should be completed in 0.3 seconds, while the multimedia application requires the same computational units with 3 seconds of time constraint. Note that the arrival rate for both application types is also the same. As a result, tasks of multimedia are completed more successfully than entertainment tasks. Similarly, rendering and image classification applications have the same time constraint and arrival rate of 1 and 20 sec/task, respectively. However, image classification requires three times computational unit than that of rendering. Therefore, rendering tasks are processed better than image classification. Moreover, even though rendering requires two times computational unit than that of the entertainment, their tasks are processed more successfully as the arrival rate is two times bigger than the entertainment. Furthermore, the maximum tolerable delay of the rendering application is three times bigger than the entertainment, which affects task completion. On the other hand, considering image classification, the number of UAVs in the environment would not affect the task success rate. The fundamental reason for this result is that image classification requires 600 computational units with 1-second delay tolerance while the capacity of a UAV is 500 computational units/sec. Therefore, none of the deployed UAVs can process image classification tasks successfully in this setting. However, since UAV deployment relieves edge server resources, it can only slightly affect the task success rate of image classification, as shown in Figure \ref{app-2}. Note that these are the results of the demonstrative scenario. AirCompSim is flexible enough to evaluate different scenarios.

\section{Conclusion and Future Work}

In this study, we proposed a new simulator, AirCompSim, in order to provide a development and research environment in which researchers can conduct experiments for the optimization of task offloading, mobility, and flying policies in an air computing environment. Considering the dynamic requirements of air computing, we developed a modular structure for the implementation so that users can extend the existing policies and event mechanism based on their needs. Moreover, we designed a scenario to represent the capabilities of our simulator. To this end, we evaluated task success rate, service time of different server types, utilization of the resources, and application-based performance based on the varying number of users and UAVs. Our results showed that AirCompSim can operate well and therefore would be used as a testbed for air computing research.

Even though the existing capabilities of AirCompSim provide many benefits, we plan to add new modules including energy calculations and also dynamically changing UAV radius based on the altitude. Thus, AirCompSim can be utilized by researchers to explore broader topics.

\bibliographystyle{IEEEtran}
\bibliography{AirComputing}

\end{document}